\documentclass[fleqn,usenatbib,floatfix]{mnras} 
\usepackage{newtxtext,newtxmath}

\usepackage[T1]{fontenc}
\usepackage{ae,aecompl}
\usepackage{graphicx} 

\usepackage{amsfonts}
\usepackage{amsmath}
\usepackage{amssymb} 
\usepackage{booktabs}
\usepackage[titletoc]{appendix}
\usepackage{aas_macros}
\usepackage[utf8]{inputenc}
\usepackage[T1]{fontenc}
\usepackage{fullpage}
\usepackage{natbib}
\usepackage[final]{pdfpages}
\usepackage{siunitx}
 
\usepackage{subcaption}

 \newcommand{\abs}[1]{\ensuremath{\lvert {#1}\rvert}}

\newcommand{\chsqr}{\ensuremath{\chi^2}}

\newcommand{\lcdm}{\ensuremath{\Lambda}CDM}

\newcommand{\DD}{\ensuremath{\mathcal{D}}}

\newcommand{\Omm}{\ensuremath{\Omega_\text{m}}}

\newcommand{\Omk}{\ensuremath{\Omega_k}}

\newcommand{\diff}{\ensuremath{\mathrm{d}}}

\newcommand{\Om}{\ensuremath{Om}}

\newcommand{\vect}[1]{\ensuremath{\mathbf{#1}}}
\newcommand{\tens}[1]{\ensuremath{\mathbfss{#1}}}

\newcommand{\be}{\begin{equation}}
\newcommand{\ee}{\end{equation}}
\newcommand{\bea}{\begin{eqnarray}}
\newcommand{\eea}{\end{eqnarray}}

\title{Model Independent Expansion History from Supernovae: 
Cosmology vs Systematics}

\author{Benjamin L'Huillier${}^1$, Arman Shafieloo${}^{1,2}$, Eric~V.~Linder${}^{1,3,4}$, Alex~G.~Kim${}^{3}$ 
\\
$^{1}$Korea Astronomy and Space Science Institute, Yuseong-gu, Daedeok-daero 776, Daejeon 34055, Korea\\
$^2$ University of  Science and Technology,  Yuseong-gu 217 Gajeong-ro, Daejeon 34113, Korea\\
$^3$ Lawrence Berkeley National Lab, 1 Cyclotron Rd., Berkeley, CA, 94720\\
$^4$ Energetic Cosmos Laboratory, Nazarbayev University, Astana, Kazakhstan
} 

\date{Accepted XXX. Received YYY; in original form ZZZ}

\pubyear{2018}

\begin{document}
\label{firstpage}
\pagerange{\pageref{firstpage}--\pageref{lastpage}}
\maketitle

\begin{abstract}
We examine the Pantheon supernovae distance data compilation in a model independent analysis to test the validity of cosmic history reconstructions beyond the concordance $\Lambda$CDM cosmology. Strong deviations are allowed by the data at $z\gtrsim1$ in the reconstructed Hubble parameter, $\ensuremath{Om}$ diagnostic, and dark energy equation of state. We explore three interpretations: 1) possibility of the true cosmology being far from $\Lambda$CDM, 2) supernovae property evolution, and 3) survey selection effects. The strong (and theoretically problematic) deviations at $z\gtrsim1$ vanish and good consistency with $\Lambda$CDM is found with a simple Malmquist-like linear correction. The adjusted data is robust against the model independent iterative smoothing reconstruction. However, we caution that while by eye the original deviation from $\Lambda$CDM is striking, $\chi^2$ tests do not show the extra linear correction parameter is statistically significant, and a model-independent Gaussian Process regression does not find significant evidence for the need for correction at high-redshifts. 
\end{abstract}

\begin{keywords}
cosmological parameters --
distance scale --
cosmology: observations --
cosmology: theory --
methods: statistical
\end{keywords}



\section{Introduction} 


Type Ia supernovae (SNIa) distance indicators have proved to be one of the most 
successful probes of the cosmic expansion history, leading to the discovery of its 
acceleration \citep{1999ApJ...517..565P,1998AJ....116.1009R}. The currently most 
complete SNIa distance 
compilation, Pantheon \citep{2018ApJ...859..101S}, has more than a thousand spectroscopically confirmed SNIa. 
Upcoming surveys such as Euclid \citep{2011arXiv1110.3193L}, Large Synoptic Survey Telescope  \citep[LSST,][]{2008arXiv0805.2366I}, or Wide-Field Infrared Survey 
Telescope  \citep[WFIRST,][]{2015arXiv150303757S} will increase the number of  
potential SNIa candidates and extend the data to higher redshifts. 
While the statistical error will improve greatly with the next generation data, it 
is important to improve our understanding of systematic uncertainties accordingly.  

This is especially true when seeking to verify or falsify the current concordance 
cosmology of \lcdm, dominated by a cosmological constant and cold dark matter. 
In order to detect robustly small variations in cosmological model, we must control 
the astrophysical and observational aspects of the data. 
Use and interpretation of SNIa survey data must  account for systematics such as the Malmquist bias, population evolution, gravitational lensing, and dust extinction. These have been reasonably well 
characterized at $z\lesssim1$ \citep{2011ApJS..192....1C, 2014ApJ...795...45S,2015ApJ...813..137R, 2018arXiv181102377B, 2018arXiv181102381H}, but are more 
uncertain at higher redshift. Some are complicated functions of the survey 
characteristics, for example  
the Malmquist bias is a well-known selection effect where 
at redshifts near the survey magnitude threshold, brighter supernovae are 
more likely to be detected, biasing the effective luminosity towards brighter 
values, therefore inferring shorter distances. 

Cosmologically, SNIa constrain the shape of the distance-redshift relation and 
hence the Hubble expansion rate that depends on the energy density contents of 
the universe. For a Friedmann-Lema\^itre-Robertson-Walker (FLRW) metric, the dimensionless 
comoving distance \DD\ is defined for any curvature parameter \Omk\ by
\begin{align}
\DD(z) & = \frac{1}{\sqrt{-\Omk}}\sin\left(\sqrt{-\Omk}\int_0^z\frac{\diff x}{h(x)}\right)\ , \label{eq:Dh} 
\intertext{where}
h^2(z) & = \Omm(1+z)^3 + \Omk(1+z)^2  \nonumber\\
&+ (1-\Omm-\Omk)\exp\left(3\int_0^z\frac{1+w(x)}{1+x}\diff x\right) \label{eq:hw}
\end{align}
is the Hubble expansion history (squared),  $\Omm$ and $\Omk$ are the matter and curvature density parameters at $z=0$, and $w(z)$ is the equation of state of dark energy. 

Given a specific cosmological model, i.e.\ $\Omm$, $\Omk$, $w(z)$, one can assess 
the goodness of fit to the data. Certainly one can test the concordance flat 
\lcdm\ model with $\Omk=0$, $w=-1$. However, it is also of interest to explore 
the cosmological data in a model independent manner, without claiming to know the 
energy density constituents and their behaviors (especially given our ignorance of 
dark energy properties). That is, we use Eq.~(\ref{eq:Dh}) in terms of the Hubble 
parameter $h(z)$, without using its specific construction in terms of components 
in Eq.~(\ref{eq:hw}). For clarity, and motivated by theoretical considerations from 
inflation and observational considerations from the cosmic microwave background (CMB), 
we do adopt spatial flatness $\Omk=0$, so $\DD=\int dz/h(z)$. We will call this 
the model independent approach, meaning it does not assume a dark energy model, 
though it does assume FLRW and flatness. 

\citet{2018PhRvD..98h3526S} reconstructed the expansion history from the Pantheon 
supernova compilation in such a model independent manner. 
The reconstructions show validity of rapidly growing expansion history at high redshifts, 
corresponding to a flattening of the luminosity distances (shorter distances than 
expected from \lcdm) or negative distance moduli relative to \lcdm. While in that 
paper the focus was on consistency with baryon acoustic oscillation (BAO) distances 
and the CMB sound horizon, exploring \DD\ vs $h$ to test the FLRW and flatness 
assumptions, and comparing growth vs expansion to test general relativity, here we 
focus on the expansion history and comparison with concordance \lcdm\ cosmology, and 
exploring implications of any deviations for cosmology, supernova properties, or 
survey systematics \citep[see also][]{2009A&A...499...21F,2018arXiv180306197T}.

In Sec.~\ref{sec:hiz} we briefly review the implications of the data on model 
independent reconstruction of cosmological quantities. We then examine three 
avenues of explanation for deviations from \lcdm: cosmology (Sec.~\ref{sec:cos}), source properties 
(Sec.~\ref{sec:sn}), and survey characteristics (Sec.~\ref{sec:malm}). We 
conclude in Sec.~\ref{sec:concl}.

\section{High Redshift Deviation} \label{sec:hiz} 

The behaviour of the smooth reconstructions in \citet{2018PhRvD..98h3526S} shows that a substantial deviation from $\Lambda$CDM is allowed by the data at high redshift. 

\begin{figure*} 
\centering 
  \includegraphics[width=\textwidth]{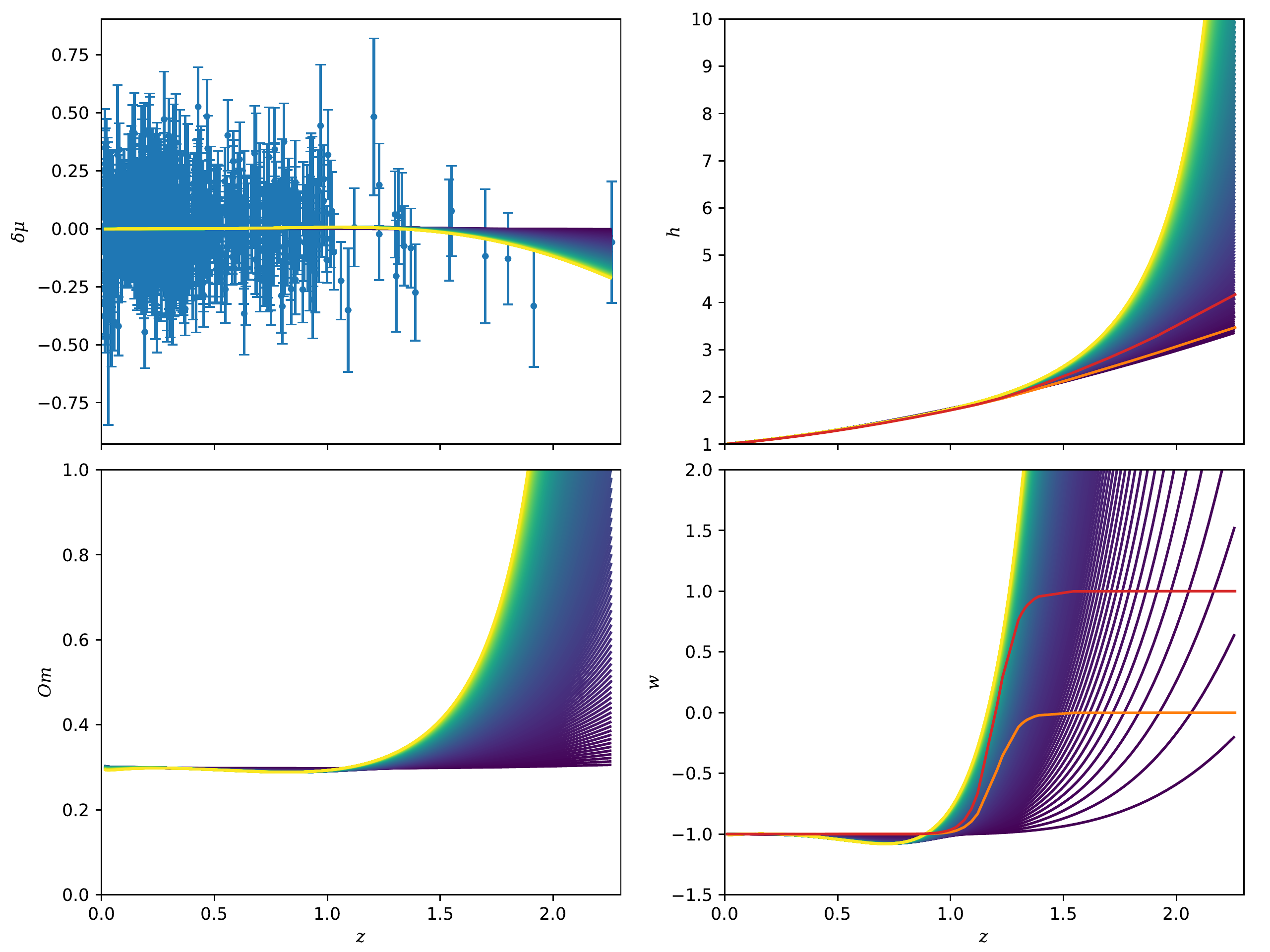} 
\caption{\label{fig:hOm_uncorr}Top left: SNIa magnitude residuals with respect to the \lcdm\ best fit. The data points 
with error bars are the residuals of the measurements, and the  coloured lines are the smooth 
reconstructions from dark blue (first iteration) to light yellow (last iteration).
All lines here give a better \chsqr\ to the data than the best-fit \lcdm. 
Top right:  dimensionless Hubble expansion parameter derived from the model independent reconstruction. 
Bottom left: \Om\ diagnostic.
Bottom right: equation of state of dark energy reconstructed assuming the best fit 
matter density $\Omm = 0.298$, corresponding to the best-fit \lcdm. 
Orange (lower) and red (upper) lines in the right panels represent two most extreme theoretical possibilities that early dark energy could behave as matter ($w=0$) or stiff matter ($w=1$). Reconstructions show much stronger deviations even with respect to these extreme models.  
}
\end{figure*}

We applied the iterative smoothing algorithm \citep{2006MNRAS.366.1081S, 2007MNRAS.380.1573S, 2017JCAP...01..015L}, taking into account the covariance matrix of the Pantheon sample \citep{2018PhRvD..98h3526S}. 
Starting from the best-fit \lcdm\ model, we stop the procedure after 200 iterations and kept all reconstructions yielding a better \chsqr\ than the best-fit \lcdm\ model, which correspond to a subsample of the reconstructions in \citet{2018PhRvD..98h3526S}.
For the sake of completeness, we briefly explain the method here. 
Starting from some initial guess for $\hat\mu_0(z)$, we iteratively obtain the reconstructed $\hat\mu_{n+1}(z)$ at iteration $n+1$ as follow:
\begin{align}
  \hat\mu_{n+1}(z) & = \hat\mu_n(z) + \frac{\vect{\delta\mu}_n^\mathrm{T} \cdot
  \tens{C}_\text{SN}^{-1} \cdot \vect{W}(z)}  
  {(1,\dots,1) \cdot \tens{C}_\text{SN}^{-1} \cdot \vect{W}(z)},
  \intertext{where the weight \vect W and residual \vect{\delta \mu_\mathrm{n}} are defined as}
  \vect{W}_i(z) & =  {\exp{\left(- \frac{\ln^2\left(\frac   {1+z}{1+z_i}\right)}{2\Delta^2}\right)}},\\
  \vect{\delta\mu}_n|_i & = \mu_i-\hat\mu_n(z_i),
\end{align} 
and $\tens{C}_\text{SN}$ is the covariance matrix of the Pantheon data. 
In case of uncorrelated data ($C_{ij} = \delta_{ij} \sigma_i^2)$), we recover the formula used in \citet{2007MNRAS.380.1573S} and \citet{2017JCAP...01..015L}. 
We thus end up with a collections of expansion histories all yielding a better \chsqr\ to the data than the best-fit \lcdm\ model, and as such, are a non-exhaustive sample of plausible expansion histories indistinguishable from \lcdm.
The top-left panel of Fig.~\ref{fig:hOm_uncorr} shows the residuals 
in blue with respect to the best-fit \lcdm\ model. 
The different smoothing iterations are shown in colour-codes, from dark blue (first iteration) to yellow (last iteration). Each curve is a viable fit to the data, 
with $\chi^2$ equal to or better than the \lcdm\ fit. We emphasize that these should be viewed as sample viable cosmologies and not a confidence region. 

The data at $z\gtrsim1$ yield negative residuals on average, pulling the reconstructed $\hat{\mu}(z)$ toward lower values with respect to the \lcdm\ best-fit. 
These low distance moduli in turn drive the corresponding sharp rise in the Hubble 
parameter $h$ seen in the upper right panel. 
The bottom-left panel shows the \Om\ diagnostic, a model independent statistic defined as \citep{2008PhRvD..78j3502S}
\begin{equation}
\Om(z)= \frac{h^2(z)-1}{(1+z)^3-1}\ .
\end{equation}
For a flat \lcdm\ universe, \Om\ is constant and equal to the matter density parameter $\Om(z) \equiv \Omm$. 
In the panel \Om\ can clearly deviate from constancy, increasingly so at high 
redshift for successive iterations. That is, the data tend to pull the smooth 
reconstructions away from \lcdm, reflecting higher $h$, lower distances, and, in 
a less model independent sense, more dark energy at higher redshifts, corresponding to 
less negative equation of state $w$.

We explore three origins for the possible deviations in the model independent  reconstructions in the following sections: 1) the data accurately reflect the underlying (non-\lcdm) cosmology and theoretical possibilities, 2) the data reflect differences in the high and low redshift supernova samples, i.e.\ some shift in source properties, and 3) the data are influenced by survey properties at high redshift such as declining detection efficiency (e.g.\ incompletely corrected Malmquist bias). 
In other words, if the data are taken at face value, then it has the following implications  for cosmology (Sec.~\ref{sec:cos}) or source properties (Sec.~\ref{sec:sn}), while Sec.~\ref{sec:malm} explore the possibility of the survival of residuals after the correction.

\section{Cosmological Interpretation} \label{sec:cos} 

The reconstructed model-independent Hubble parameter closely traces the $\Lambda$CDM behavior 
out to $z\approx1$ and then it can rapidly increase above it at higher redshift. 
This implies an extra energy density relative to \lcdm\ at high redshift, which can be interpreted as a dark 
energy that fades away less strongly than a cosmological constant. In order to preserve the 
\lcdm\ behavior at $z\lesssim1$, the dark energy equation of state has to be near $w\approx-1$ 
at low redshift before becoming less negative at early times. 

The apparent 
severity of the cosmology shift is perhaps more clearly appreciated if we relax model 
independence to the extent of defining a separate matter component with $\Omm$ given by 
the best fit \lcdm\ 
and plotting the effective dark energy equation of 
state $w(z)$ in the lower right panel of Fig.~\ref{fig:hOm_uncorr}. 
We can compute the equation of state behavior $w(z)$ from the Hubble parameter by 
\be 
w(z)=-1+\frac{1}{3}\frac{d\ln[h^2-\Omm(1+z)^3]}{d\ln(1+z)} \ . 
\ee 
Clearly the results show that this is not a simple evolution as for  a freezing quintessence, or even a transition from an early dark energy scaling as matter.  

Such $w(z)$ behavior does not correspond to any standard dark energy model. 
A sharp transition is required at $z\approx1$ from nearly a cosmological constant 
behavior ($w=-1$) to large positive values of $w$ (this is independent of the exact 
value of $\Omm$ taken). Even phenomenological models with a rapid  (much faster than Hubble time) transition to 
early dark energy matter behavior \citep[$w=0$,][]{2018MNRAS.473.2760S}, or more extremely stiff matter behavior 
 \citep[$w=1$,][]{1972MNRAS.160P...1Z}, do not provide close fits. 
Moreover, any such model that approaches $w\approx0$ by $z\approx1.5$ would unviably 
alter the growth of structure and the integrated Sachs-Wolfe effect in the cosmic microwave background.

A purely cosmological interpretation of the high-redshift behaviour carries grave challenges.
Such extreme expansion histories require dark energy models that would violate matter domination, and thus do not seem viable overall. 
Therefore, rather than cosmology, we look for systematics in the data and consider in the next section source and survey characteristics as origins of the model-independent deviation from \lcdm.

\begin{figure*} 
\includegraphics[width=\textwidth]{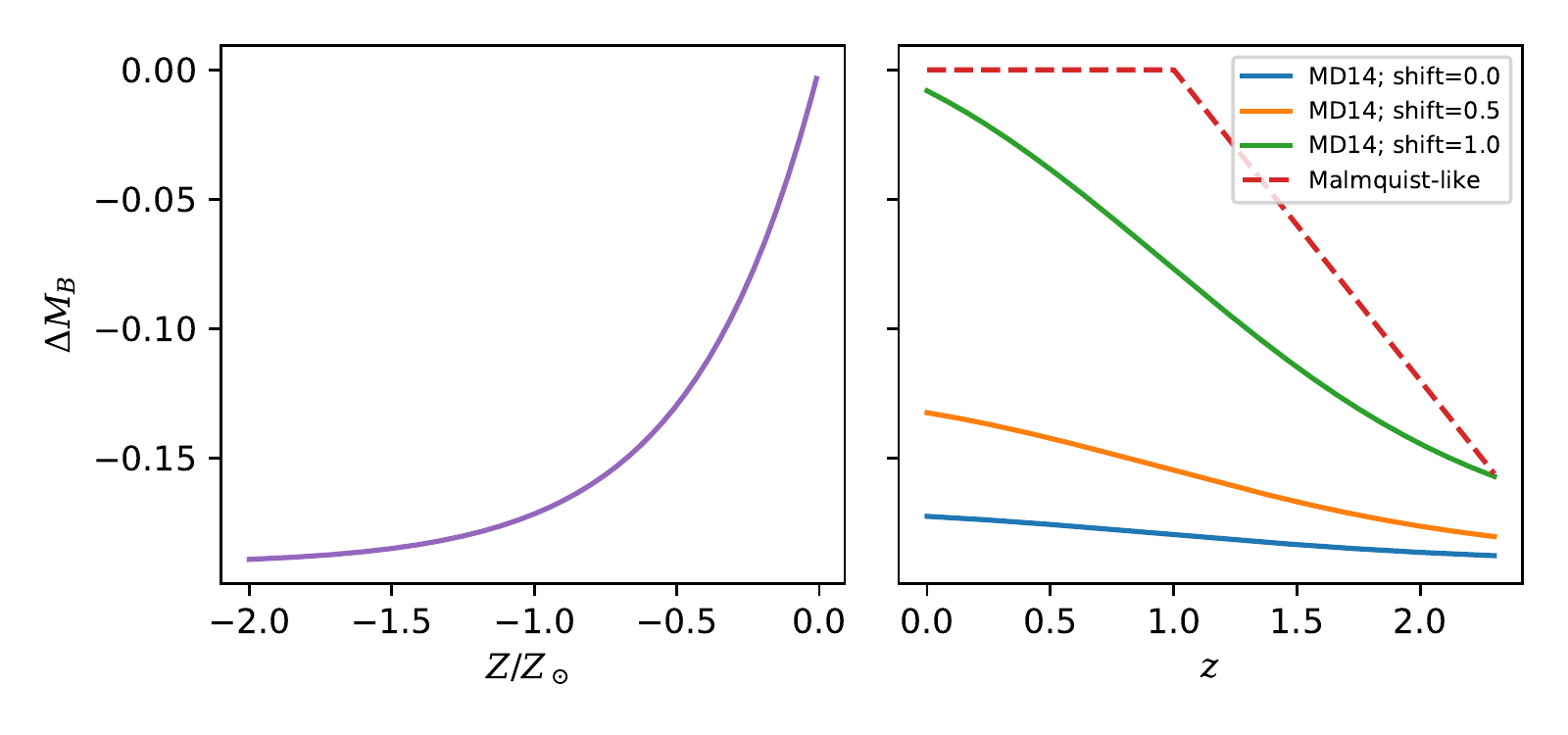}
\caption{\label{fig:Mz}Magnitude bias $\Delta M_B$ as a function of metallicity (left) and redshift (right). Depending on the $Z(z=0)$ value, the magnitude bias redshift evolution can be mild or significant, as shown for three cases in the right panel. 
The red dashed line is the empirical correction Eq.~(\ref{eq:lincorr}) from \S\ref{sec:malm}. Note that only the magnitude difference with redshift, not the absolute value, matters for cosmology. 
} 
\end{figure*}

There is one other cosmological source for the deviation worth considering first. Gravitational 
lensing changes the magnitudes of sources, increasingly so for higher redshift sources. 
While for a large number of sources at a given redshift the apparent distance is unaffected 
at linear order, for a small number of sources they will be preferentially brightened as a selection effect. 
The average dispersion effect is corrected for in the Pantheon compilation. Let us briefly 
consider how much the correction would need to be off to undo the deviation from \lcdm\ in 
Fig.~\ref{fig:hOm_uncorr}. 

The 
Pantheon analysis assumes the form $\sigma_{\rm lens}=0.055z$ from 
\citet{2010MNRAS.405..535J}. If instead we took the form $0.093z$ from 
\citet{2005ApJ...631..678H}, or the form more complete at higher redshift, 
$0.1z/(1+0.07z)$ from \citet{2007APh....27..213A}, then we would expect an 
uncorrected deviation of 0.053 at $z=1.5$. Interestingly, we see in Sec.~\ref{sec:malm} that the 
deviation from \lcdm\ corresponds to a deviation of roughly 0.06 at $z=1.5$. 
We do not claim this is the cause of the deviation -- every source would need to be 
magnified at the $\sim1\sigma$ level -- but we urge caution in treating $z>1$ distance 
indicators without a robust lensing probability distribution function, and look 
forward to surveys that will put sufficient $z=1$--1.7 SNIa on the Hubble diagram to 
allow for safety in numbers from lensing.

\section{Supernova Property Interpretation} \label{sec:sn} 

In this section, we focus on supernovae or environmental effects that might bias the absolute magnitude through 
drift of the observed supernova population. One possibility 
is the metallicity of the host galaxy. 
\citet{2016ApJ...818L..19M} found this could affect the peak magnitude of a supernova as 
\begin{align}
\label{eq:MZ}
\Delta M_\mathrm{B} & = -2.5 \log_{10}\left(1-0.18 \dfrac{Z}{Z_\odot}\left(1-0.10 \dfrac{Z}{Z_\odot}\right)\right)-0.191\,, 
\end{align} 
where $Z_\odot$ is solar metallicity. 
The magnitude bias as a function of the host metallicity is shown in the left-hand panel of Fig.~\ref{fig:Mz}.

A constant offset would not affect supernova cosmology, but an unaccounted for  
redshift evolution could. The cosmic mean metallicity $Z_b$ as a function of redshift is given by \citep{2014ARA&A..52..415M}
\begin{align}
Z_b(z) & = y \frac{\rho_*}{\Omega_\mathrm{b}\rho_{0\mathrm{c}}}Z_\odot\ , \label{eq:Zz}
\intertext{where}
\rho_*(z) & = (1-R)\int_z^\infty \psi \frac {\diff z'}{H(z')(1+z')}\ , \\
\rho_{0\text{c}} & = \frac{3H_0^2}{8\pi G}
\intertext{is the critical density of the Universe, and the star formation rate is }
\psi(z) & = 0.015\frac{(1+z)^{2.7}}{1+[(1+z)/2.9]^{5.6}} \si{M_\odot.yr^{-1}.Mpc^{-3}}.
\end{align}
The yield $y$ and the return rate $R$ depend on the initial mass function, and for a Salpeter IMF, are about $y=0.02$ and $R=0.27$.

Combining equations~\eqref{eq:MZ} and~\eqref{eq:Zz}, we obtain the absolute magnitude offset as a function of redshift, shown in the right-hand panel of Fig.~\ref{fig:Mz}. 
Following \citet{2016Natur.534..512B}, we shifted the relation by \{0,+0.5,+1\} dex in blue (lower), orange (middle), and green (upper), accounting for the uncertainty in the redshift zero metallicity expression in 
comparison with data. 

None of these potential residual corrections match the shape of the 
empirical correction Eq.~(\ref{eq:lincorr}) of the next 
section. 
However, we report in Table~\ref{tab:chi2Om} the impact of such metallicity corrections 
on $\Omm$ and their associated $\Delta\chsqr$ with respect to the \lcdm\ uncorrected 
best-fit. 
It is worth noting that none of these corrections yield a statistically significant improvement to \chsqr. 
In addition, note the case with the shift of +1 dex, which is closest in 
magnitude to the empirical correction of the next section, yields a best-fit $\Omm = 0.251$, which is significantly lower. 
Therefore, these considerations do not seem to favour unaccounted-for metallicity evolution as the cause for the blowup of the model-independent reconstructions of $h$ at $z\geq 1$.

\begin{table}
\centering
\caption{\label{tab:chi2Om}$\Delta\chsqr$ and \Omm\ after  correction for a potential  magnitude bias due to the host galaxy metallicity evolution of Eq.~\eqref{eq:Zz}. }
\begin{tabular}{lcc}
\toprule
& $\Delta\chsqr$ & \Omm \\
\midrule
\lcdm & 0 &  0.298\\
$\Delta M_B(z)$ correction & -0.040 & 0.293\\
$\Delta M_B(z)$ correction+0.5 & -0.109 & 0.282\\
$\Delta M_B(z)$ correction+1 & -0.176 & 0.251\\
Correction \eqref{eq:lincorr} & -0.902 & 0.293\\
\bottomrule
\end{tabular}
\end{table}


\section{Survey Property Interpretation} \label{sec:malm}

Survey selection effects also can affect the Hubble diagram of the 
distance-redshift relation, and the cosmological quantities derived 
from it. The most well known example is Malmquist bias, where supernovae 
near the upper redshift limit of the survey have a preferential selection effect for the intrinsically 
brighter members of the population due to 
the survey flux and signal to noise limits. Other possible effects include 
 filter zeropoints and color corrections, though we note the Pantheon compilation 
has put great effort into calibration \citep{2018ApJ...857...51J,2018ApJ...859..101S}. 
Recent techniques for dealing with sample selection generally are 
discussed in \cite{2017arXiv170603856H,2018arXiv181102381H}. 

\begin{figure}
  \centering
  \includegraphics[width=\columnwidth]{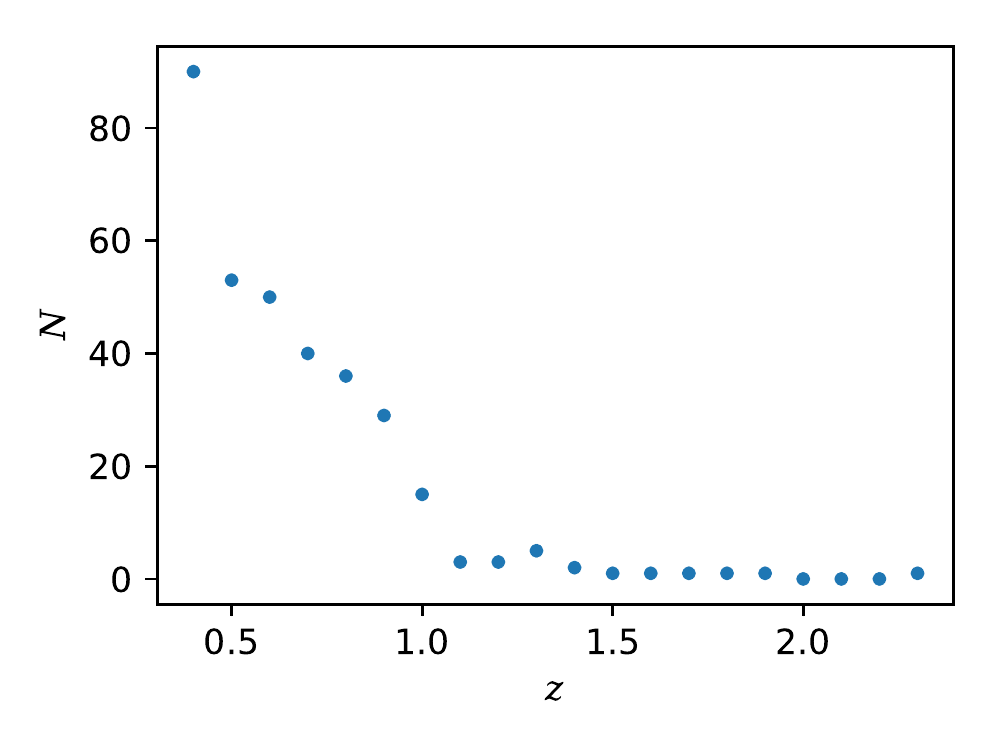}
  \caption{\label{fig:density}%
    Density of data points, binned in redshift for purely visual purposes. 
  }
\end{figure}

To mock up the effect of a high redshift selection bias (which we do not suggest is a Malmquist bias, simply that high redshift is the most likely place for bias), we consider 
a simple linear trend above a certain redshift, where the selection 
starts to bias the magnitude. 
In other words, we are considering the possibility of residuals after the current best efforts at correction.
To choose the transition redshift we examine 
the data. 
Figure~\ref{fig:density} shows the data density as a function of redshift; one might expect that selection effects are increased by a 
sparse sampling with just a few supernovae per redshift interval. 
The data density drops to low values for $z\gtrsim1$, with just a  few points per interval of $\Delta z=0.1$ for $z=1$--2. 
In addition, in the lower-left panel of Fig.~\ref{fig:hOm_uncorr}, $\Om$ is consistent with a constant level at low-$z$, and starts to depart from constant after $z\gtrsim 1$, suggesting that any new element enters only there.
Therefore, we apply the following empirical correction at  $z>z_c=1$ (corresponding to 23 points): 

\[  
\mu  \rightarrow \mu+\delta\mu\ , 
\]  
where 
\be 
\delta\mu   =   \label{eq:lincorr} 
  \begin{cases}	
A(z-z_c),& {\rm if\ } z>z_c\\ 
0, & {\rm otherwise,} 
  \end{cases} 
\ee 
and where $A$ is a free parameter to fit. 

\begin{figure}
\includegraphics[width=\columnwidth]{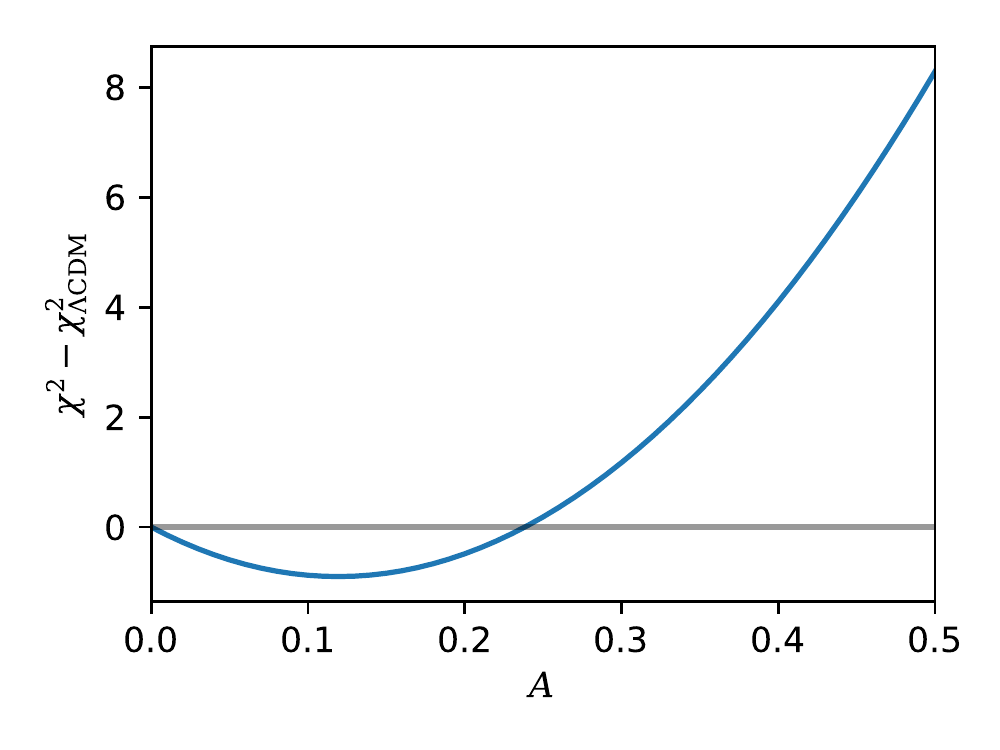}
\caption{\label{fig:contours}%
$\Delta\chi^2$ as a function of high redshift magnitude correction factor $A$. 
}
\end{figure}

Figure~\ref{fig:contours} shows $\Delta\chi^2 = \chi^2-\chi^2_0$, where
$\chi^2_0$ is the $\chi^2$ of the best-fit flat-\lcdm\
model to the uncorrected Pantheon data.  
The best $\chi^2$ is located at $A=0.12$ with $\Delta\chi^2 = -0.9$, which is  not statistically significant for one added degree of freedom (dof). 
Since the correction only affects the last 23 data points, it has little impact on the best-fit cosmology (\Omm = 0.293 versus 0.298).   
In the following, we use this value of $A=0.12$ to ``correct" the data.

\bigskip

\begin{figure*} 
\centering 
  \includegraphics[width=\textwidth]{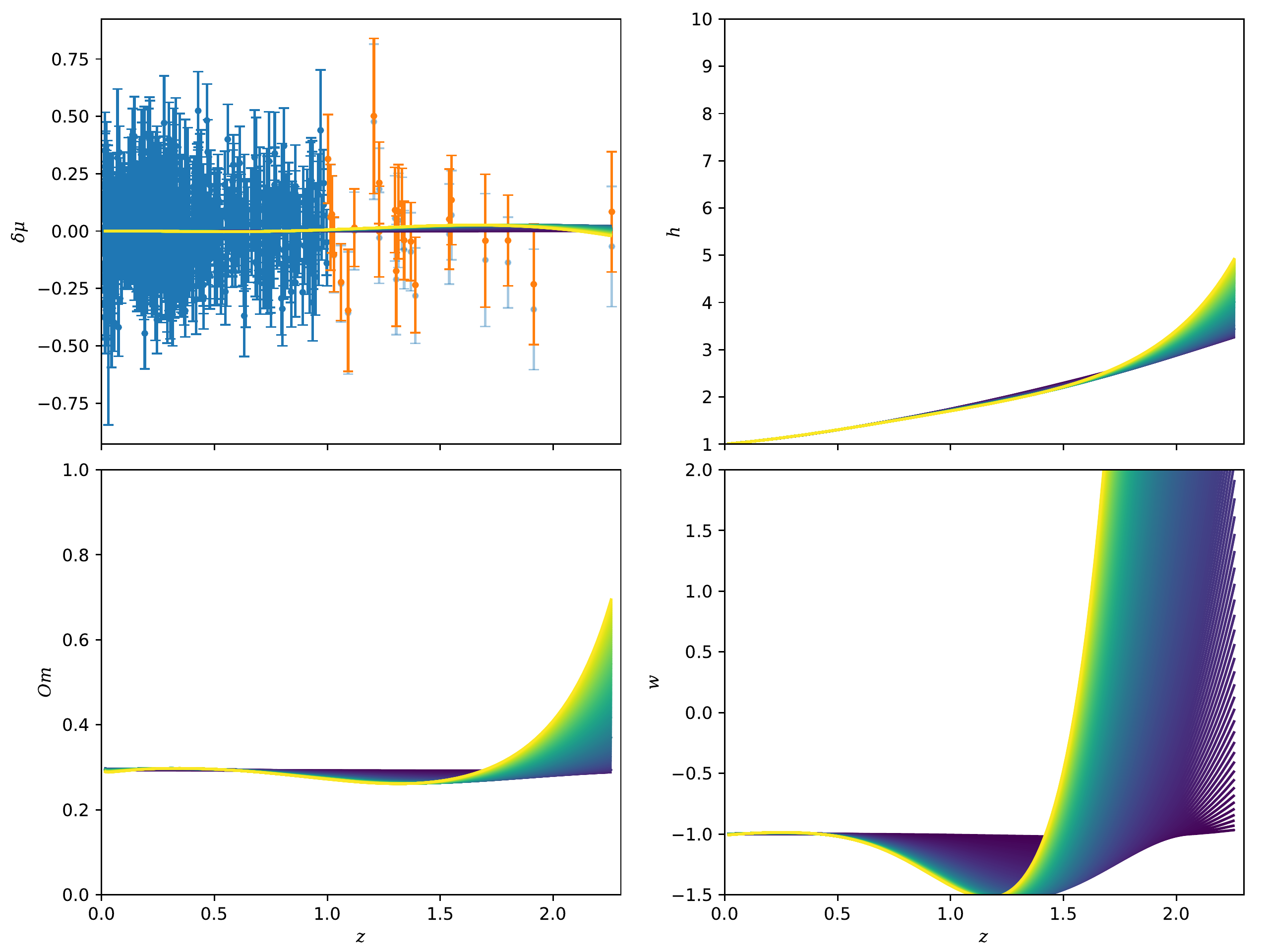} 
\caption{\label{fig:hOm_corr}Top left: residuals with respect to the \lcdm\ best fit. The data points 
with error bars are the residuals of the SNIa measurements (with orange points including the high redshift correction), and the  coloured lines are the smooth 
reconstructions from dark blue (first iteration) to yellow (last iteration), using the corrected data. 
Top right:  dimensionless Hubble parameter. 
Bottom left: \Om\ diagnostic.
Bottom right: equation of state of dark energy for the best-fit $\Omm = 0.293$.
}
\end{figure*}

After finding the best-fit flat \lcdm\ cosmology for the $\mu$ corrected according to 
Eq.~\eqref{eq:lincorr}, we implement the smoothing method to the corrected data using the best-fit value for $A$ 
to obtain the cosmological quantities. 
Figure~\ref{fig:hOm_corr} shows the same plots as Fig.~\ref{fig:hOm_uncorr}, with the corrected data (for $A=0.12)$. 
The corrected data points are shown in orange in the top-left panel. 
In the other three panels,  
the reconstructions then show tighter agreement in behavior out to higher redshift, and do not lead to a blow-up in $h$ or \Om\ for almost all the redshift range. 
We caution that the dark energy equation of state $w$, 
being a second derivative of the data, is always more 
sensitive to statistical fluctuations.

\bigskip

To quantify the significance of the deviation at high-redshift, we use the crossing statistics \citep{2011JCAP...08..017S,2012JCAP...05..024S}.
The 0th  mode of the statistics is 
\begin{align}
T_0 & = \left(\sum_{i=1}^N \frac{\mu_{\text{model},i}- \mu_{\mathrm{data},i}}{\sigma_i}\right)^2,
\end{align}
which is related to the pull, while the $N$th mode is the \chsqr. 
Table~\ref{tab:T0chi2} shows $T_0$ and \chsqr\ for the 23 points with $z>z_\text{c}$.

\begin{table} 
\centering
\caption{\label{tab:T0chi2}Crossing statistic mode 0  ($T_0$) and mode N (\chsqr), for the 23 data points with $z>z_\text{c}$, for different corrected (c) and uncorrected (uc) parametric (\lcdm) and non-parametric (smoothing) ``theory" values,  and data. 
Italics indicates that the fitting is run with the opposite set (corrected vs uncorrected) as the data, and should be used only as a crosscheck. 
}
\begin{tabular}{lllll}
\toprule
& \multicolumn{2}{l}{Data-uc} & \multicolumn{2}{l}{Data-c} \\
& $T_0$ & $\chsqr$ & $T_0$ & \chsqr \\
\midrule
 \lcdm-uc & 6.94 & 16.35  & {\it 3.47} & {\it 15.40}\\
 \lcdm-c & {\it 11.33} & {\it 16.48}   & 1.27 & 15.35 \\
 smoothing-uc & 1.06  & 15.60  & {\it 12.03} & {\it 16.15} \\
 smoothing-c & {\it 25.12} & {\it 16.95} & 0.26 & 15.45 \\
\bottomrule
\end{tabular}
\end{table}

The smoothing applied to the corrected data yields a similar \chsqr, but its associated $T_0$ is much lower: the data are more evenly distributed around 0. (A useful crosscheck is that the uncorrected data prefer the uncorrected smoothing, and the corrected data prefer the corrected smoothing.) 
To assess the significance of the obtained $T_0$, we simulated 1000 realizations of a Gaussian random variable  $\sim \mathcal{N}(0,1)$, and measured the distribution of $T_0$. 
The expected $1\sigma$ of the distribution of $T_0$ is about 32. 
Therefore, all values reported here are within $1\sigma$.  
We note that the \chsqr\ for the smoothing of the corrected data (15.45) is slightly larger than for the \lcdm\ fit to the same data (15.35). 
The reason is that the smoothing algorithm considers the total \chsqr\ to the whole data (to improve the fits iteratively), while here we are just comparing the \chsqr\ for the last 23 data points. 
The correction~\eqref{eq:lincorr} has in fact made \lcdm\ such a good fit to the data at $z>1$ that smoothing is not really improving much in this range.

As a further, independent test to assess whether the data prefer any correction, we also applied Gaussian Processes to the residuals. 
The marginal likelihood does not have a clear peak for the correlation length and does not prefer a nonzero amplitude, meaning that the data are consistent with \lcdm\ and do not suggest need for a Malmquist-like correction despite the visual impact of Fig.~\ref{fig:hOm_uncorr} (see Appendix~\ref{sec:GP}).

\section{Summary and Conclusion} \label{sec:concl} 

Model independent reconstructions of the expansion history from the Pantheon supernovae compilation allow substantial deviation from the standard $\Lambda$CDM model at redshifts greater than 1. We studied three possibilities for these deviations. 

For cosmology, the most radical theoretical possibilities, that dark energy can behave as matter or stiff matter at early stages, do not behave in as extreme a manner as cases of the reconstructions. Similarly, while gravitational lensing can cause apparent cosmology deviations, the magnitude of the apparent effect seems too large for this, though better knowledge of the lensing distribution function is needed. This makes it highly unlikely that the cause of the extreme deviations from 
the $\Lambda$CDM model should be interpreted as cosmological. 

As for source properties, we studied the possible effect of the host galaxy metallicity redshift evolution on the peak magnitude of the type Ia supernovae and their effect on the reconstructions. We derived the magnitude bias - redshift relation, and investigated its dependence on the still incompletely determined zero redshift offset. This mechanism also could not reproduce the reconstructions at high redshifts while preserving the low redshift cosmology, though it could have more modest effects. 

Regarding survey characteristics, we explored an empirical high-redshift correction of the distance moduli in the Pantheon compilation. Such selection effects, such as Malmquist bias, are often difficult to fully correct the data for. Using a simple linear correction for $z> z_c=1$, we showed the blow-up of the expansion history at high redshifts could be strongly reduced. While visually the effect of the corrections on the reconstructions are substantial, the improvement on the $\chi^2$ fit (around $0.9$) is marginal. Furthermore, using Gaussian Process regression as well as crossing statistics, we examined the significance of the improvement. 

Our results show that the correction is not required statistically: current Pantheon data is 
consistent with \lcdm\ despite significant deviations also being allowed in a model independent expansion history. We eagerly look forward to more data at $z\gtrsim1$ that would have greater leverage on confirming \lcdm\ -- or pointing toward significant effects in cosmology, gravitational lensing, source property evolution, or survey selection effects.


\section*{Acknowledgements} 
This work benefited from the Supercomputing Center/Korea Institute of Science and Technology Information with supercomputing resources including technical support (KSC-2016-C2-0035 and KSC-2017-C2-0021) and the high performance computing clusters Polaris and Seondeok at the Korea Astronomy and Space Science Institute. A.S. would like to acknowledge the support of the National Research Foundation of Korea (NRF- 2016R1C1B2016478). A.S. would like to acknowledge the support of the Korea Institute for Advanced Study (KIAS) grant funded by the
Korea government. EL is supported in part by the Energetic Cosmos Laboratory and by the U.S.\  Department of Energy, Office of Science, Office of High Energy Physics, under Award DE-SC-0007867 and contract no.\ DE-AC02-05CH11231.




\bibliographystyle{mnras}
\bibliography{biblio} 



\appendix

\section{GP regression}

\label{sec:GP}

\begin{figure*}
  \centering
  \includegraphics[width=\textwidth]{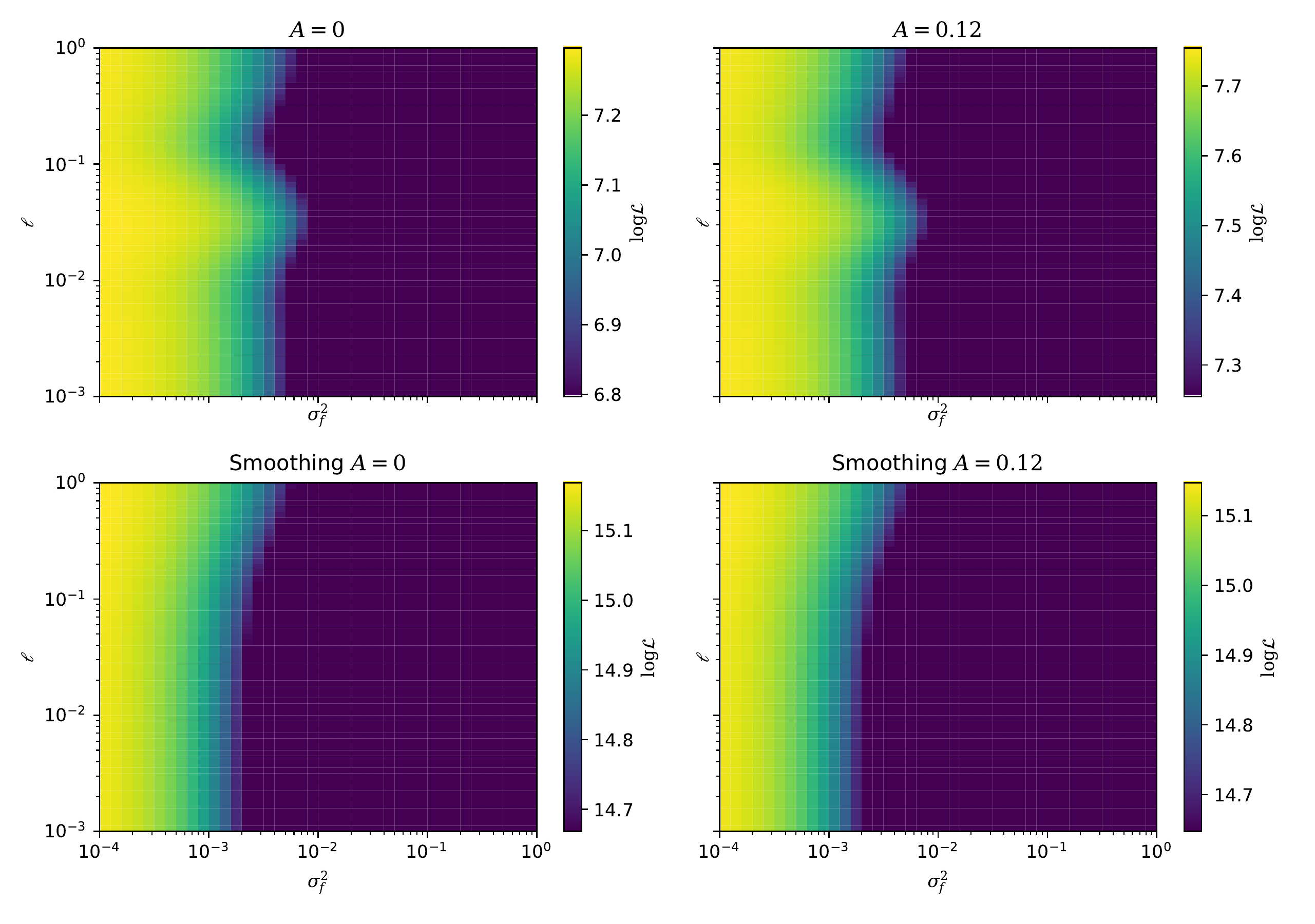}
  \caption{\label{fig:GP}Log Marginal Likelihood of the hyperparameters $(\sigma_f^2,\ell)$ for the uncorrected (left) and corrected (right, $A=0.12$) cases, without (top) and with (bottom) smoothing. No significant effect (hence small $\sigma_f^2$) is preferred.}
\end{figure*}

We applied Gaussian Processes (GP) regression \citep{2006gpml.book.....R} to the residuals (with
respect to the best-fit, uncorrected \lcdm\ model) at $z>z_c$.
GP have been widely used in the community to reconstruct the expansion history, the equation of state of dark energy, the cosmic growth rate, and more \citep[see, e.g.,][]{2010PhRvD..82j3502H,2010PhRvL.105x1302H,2011PhRvD..84h3501H,2012PhRvD..85l3530S,2013PhRvD..87b3520S}.

Starting from a training set of points $(\vect{x},\vect y = f(\vect x)+\vect \varepsilon)$ where $\varepsilon$ is a Gaussian noise with mean zero and covariance \tens{C}, and \vect{y} are the measured values, 
we can model $f$ as a stochastic process with covariance \tens K, and reconstruct $\vect{f}_*(\vect{x}_*)$ at the test points $x_*$.

The joint-distribution of the training outputs \vect{y} and the test output $\vect{f}_*$ is a Gaussian distribution given by
\begin{align}
  \begin{bmatrix}
	\vect{y} \\
    \vect{f}_*
  \end{bmatrix} & 
  	\sim \mathcal{N}\left(
  	\vect{0},
    \begin{bmatrix}
  		\tens{K}(X,X) + \tens{C} & \tens{K}(X,X_*)\\
  		\tens{K}(X_*,X)          & \tens{K}(X_*,X_*)
  	\end{bmatrix}
  \right)
\end{align}
where \tens{C} is the covariance of the data.

For a given kernel, the covariance between pairs of random variables \vect{u} and \vect{v} is thus given by$\tens{K}(f(\vect{u}),f(\vect{v})) = k(\vect{u},\vect{v})$, where $k(\vect{u},\vect{v})$ is the covariance kernel. 
We use the squared exponential kernel defined as 
\begin{align}
k(\vect{u},\vect{v};\sigma_f,\ell) & = \sigma_f^2\exp{\left(-\frac{\abs{\vect{u}-\vect{v}}^2}{2\ell^2}\right)}\ ,
\end{align}
where $(\sigma_f,\ell)$ are two hyperparameters controlling the amplitude and the correlation scale.

The log marginal likelihood (LML) is given by
\begin{align}
\log p(\vect y|X) & = -\dfrac{1}{2} \vect{y}^T(\tens K+\tens C)^{-1}\vect y -\dfrac{1}{2}\log\abs{\tens K+\tens C} - \dfrac{n}{2}\log2\pi.
\end{align}
A clear preference for $\sigma_f>0$ would mean that the data  suggest that a correction to the mean function (i.e.\ \lcdm) is needed. 
Fig.~\ref{fig:GP} shows the log marginal likelihood (LML) of the
hyperparameters $(\sigma_f^2,\ell)$. 
In all cases of the uncorrected and corrected, and unsmoothed and smoothed data, the shapes of the LML are similar, with the corrected case having a slightly higher likelihood. 
It is maximal at low values of $\sigma_f$, i.e.\  consistent with no correction, suggesting no deviation from the mean function in both corrected and uncorrected cases. 

\label{lastpage}
\end{document}